\documentclass[12pt]{article}
\usepackage{amssymb}
\usepackage{amsmath}
\usepackage{graphicx}
\usepackage{bm}
\newcommand{\mathsym}[1]{{}}

\usepackage{graphicx}
\usepackage{rotating}
\usepackage{color}
\usepackage{cancel}

\newcommand{\bra}{\begin{array}}
\newcommand{\era}{\end{array}}
\newcommand{\beq}{\begin{equation}}
\newcommand{\eeq}{\end{equation}}
\newcommand{\beqar}{\begin{eqnarray}}
\newcommand{\eeqar}{\end{eqnarray}}

\newcommand{\be}{\begin{equation}}
\newcommand{\ee}{\end{equation}}
\newcommand{\bea}{\begin{eqnarray}}
\newcommand{\eea}{\end{eqnarray}}
\newcommand{\bd}{\begin{displaymath}}
\newcommand{\ed}{\end{displaymath}}

\newcommand{\h}{\hbar}

\newcommand{\lb }{ \left (}
\newcommand{\rb }{ \right )}

\newcommand{\p }{ \partial}

\begin{document}

\vspace{20pt}

\begin{center}

{\LARGE \bf New generalized uncertainty principle from the doubly special relativity

\medskip
 }
\vspace{15pt}

{\large  Won Sang Chung${}^{1,\dag}$ and Hassan Hassanabadi${}^{2,\ddag}$
}

\vspace{15pt}
{\sl ${}^{1}$Department of Physics and Research Institute of Natural Science,\\
 College of Natural Science,\\
 Gyeongsang National University, Jinju 660-701, Korea}\\

{\sl ${}^{2}$ Faculty of physics, Shahrood University of Technology, Shahrood, Iran }

\vspace{5pt}
E-mails:  {$
{}^{\dag}$mimip44@naver.com}
{${}^{\ddag}$ h.hasanabadi@shahroodut.ac.ir}

\vspace{10pt}
\end{center}

\begin{abstract}

Based on the doubly special relativity we find a new type of generalized uncertainty principle (GUP) where the coordinate remain unaltered at the high energy while the momentum is deformed at the high energy so that it may be bounded from the above. For this GUP, we discuss some quantum mechanical problems in one dimension such as box problem,  momentum wave function, and harmonic oscillator problem.


\end{abstract}

 \today

\section{Introduction}

As one method for curing some problems in quantum gravity, the generalization of the uncertainty relation has come, which is called a generalized uncertainty principle (GUP) [1-35]. There has been much development for the GUP formulation and GUP-corrected quantum systems. The generalized uncertainty principle (GUP) is given by the modified commutation relation
\be
[ X, P] = i ( 1 + \beta P^2),
\ee
where $X$ and $P$ is the position operator and the momentum operator, respectively, and $\beta$ is a small parameter given by $\beta= \beta_0/M_{pl} c^2 $, and  $M_{pl}$is the Planck mass and $\beta_0$ is
of order the unity, and we set $\h=1$. The GUP guarantees the non-zero minimal length and is related to the quantization of gravity.

More generally, the modified commutation relation can be written as [23,24,25,33,34,35]
\be
[ X, P] = i F(P),
\ee
where $F(P)$ is called a GUP deformation function which reduces to $1$ when the GUP effect is ignored. From now on we will call the eq.(2) a generalized GUP. Here, $(X, P)$ implies the coordinate and momentum at the high energy while $(x, p)$ is the coordinate and momentum at the low energy where $x$ and $p$ is defined through $[x, p] = i$. At the high energy the coordinate and momentum is assumed to be  deformed through
\be
x \rightarrow X=x , ~~~ p \rightarrow P= f(p)
\ee
The eq.(3) implies that the position remains undeformed at the high energy while the momentum  is deformed at the high energy. For the GUP (1), we have
\be
P = \frac{1}{\sqrt{\beta}} \tan ( \sqrt{\beta p})
\ee
In this case the momentum at the high energy is not bounded, $ - \infty < P < \infty$. This seems strange because the momentum at the high energy should be bounded if we consider the doubly special relativity (DSR) [36-40]. Indeed DSR says that the momentum has the maximum called a Planck momentum which is another invariant in DSR.

Therefore, in order to construct the new  GUP with both minimum length and maximum momentum, we should find the  mapping ( or deformation)  $  X=x,  P= f(p)$ with $ f(\pm \infty) = \pm \kappa
 $. Thus, from this map, the momentum operator has the maximum value ( Planck momentum $\kappa$ ).

In this paper we are to find a new type of GUP  where the coordinate remain unaltered at the high energy while the momentum is deformed at the high energy so that it may be bounded from the above. Our GUP model comes from the concept of DSR. This paper is organized as follows:
In section 2 we discuss the brief review of representation of generalized GUP. In section 3 we discuss the
new GUP from the concept of DSR. In section 4 we discuss the momentum wave function in a position representation. In section 5 we discuss one dimensional box problem. In section 6 we discuss harmonic oscillator problem.

\section{ Brief review of representation of generalized GUP}

Now let us reconsider the generalized GUP
\be
[ X, P] = i F(P),
\ee
For this commutation relation we have two representations.

\subsection{ Deformed momentum representation}
The deformed momentum representation for the algebra (5) is
\be
X = i  F(p) \frac{\p}{\p p}
, ~~~
P = p
\ee
The momentum representation acts on the square integrable functions $\Phi(p) \in {\cal L}^2 \lb -A, A; \frac{dp}{F(p)}\rb$  with $\phi(\pm A) =0$ and
the norm of $\phi $ is given by
\be
|| \Phi ||^2 = \int_{-\infty}^{\infty}  \frac{ dp}{F(p)} |\Phi (p)|^2
\ee
For the standard GUP (1) we have $A = \infty$.
The Schr\"odinger equation reads
\be
\left[ \frac{p^2}{2m} + V\lb i  F(p) \frac{\p}{\p p} \rb  \right] \Phi(p) = E\Phi(p)
\ee

\subsection{ Position representation}
The position representation for the algebra (5) is
\be
X = x, ~~~~
P = f(p) = f \lb \frac{1}{i} \p_x \rb
\ee
where the function $f$ is obtained from
\be
p = \int^P \frac{dP}{ F(P)}
\ee
The position representation acts on the square integrable functions $\psi(x) \in {\cal L}^2 \lb -\infty , \infty; dx \rb  $  and
the norm of $\psi $ is given by
\be
|| \psi ||^2 = \int_{-\infty}^{\infty}  dx  |\psi (x)|^2
\ee
The Schr\"odinger equation reads
\be
\left[ \frac{1}{2m} \lb f \lb \frac{1}{i} \p_x \rb\rb^2  + V\lb x\rb  \right] \psi(x) = E\psi(x)
\ee

\subsection{ Undeformed momentum representation}
The undeformed momentum representation for the algebra (5) is obtained from the position representation with replacing $x = i \p_p$, as
\be
X = x = i \p_p, ~~~~
P = f(p) = f \lb p \rb
\ee
The undeformed momentum   representation acts on the square integrable functions $\phi(p) \in {\cal L}^2 \lb -\infty , \infty; dp \rb  $  and
the norm of $\psi $ is given by
\be
|| \phi ||^2 = \int_{-\infty}^{\infty}  dp  |\phi (p)|^2
\ee
The Schr\"odinger equation reads
\be
\left[ \frac{1}{2m} \lb f (p)  \rb^2  + V\lb   i \p_p \rb\right] \phi(p) = E\phi(p)
\ee

\section{ New GUP from the concept of DSR}

In the DSR, the energy-momentum relation is deformed into
\be
E^2 = p_0^2 + m^2 + h ( |p_0|, \kappa),
\ee
where we demand that
\be
\lim_{ \kappa \rightarrow \infty} h ( |p_0|, \kappa) =0
\ee
and $p_0$ means that it is not a operator but a number, and we set $c=1$. The undeformed momentum $p_0$ ( momentum at the low energy) can be related to the deformed momentum $P_0$ ( momentum at the high energy) as follows:
\be
P_0 = f ( p_0)
\ee
If we choose
\be
P_0 =f ( p_0) = \frac{p_0}{ 1 + \frac{|p_0|}{\kappa}},
\ee
we have the modified dispersion relation
\be
E^2 = P_0^2 + m^2 = \left[ \frac{p_0}{ 1 + \frac{|p_0|}{\kappa} }\right]^2 + m^2
\ee
or
\be
E^2 = p_0^2 + m^2  - 2 \frac{|p_0|}{\kappa} p^2 + 3 \frac{p^4}{\kappa^2 } + \cdots
\ee
We can easily check that the choice (19) obeys
\be
\lim_{p_0 \rightarrow \pm \infty} P_0 = \lim_{p_0 \rightarrow \pm \infty} f(p_0) = \pm \kappa
\ee
which gives $ |P_0|\le \kappa$.

Based on the DSR, we consider the following relation for the momentum operators 
\be
P = \frac{p}{ 1 + \frac{|p|}{\kappa}},
\ee
where $|p|$ is the magnitude of undeformed momentum operator $p$, or $|p| = \sqrt{ p^2}$.
The inverse transformation is
\be
p =  \frac{P}{ 1 - \frac{|P|}{\kappa}}
\ee
From the requirement $|p|\ge 0$ we get $|P| \le \kappa$ which gives the upper bound for the momentum at the high energy. The limit $p\rightarrow \pm \infty$ corresponds to $P =\pm \kappa$, which implies that there exists the maximum momentum ( Planck momentum ) in our model. The eq.(23) gives the following commutation relation
\be
[ X, P] = i \lb 1 - \frac{|P|}{\kappa}\rb^2,
\ee
which gives
\be
\Delta X \Delta P \ge \frac{1}{2} \lb 1 - 2 \frac{\langle |P|\rangle}{\kappa} + \frac{1}{\kappa^2} (\Delta P)^2 \rb,
\ee
where we set $\langle P \rangle =0$. From now on we call the  above GUP the DSR-GUP.
The eq.(26) gives the minimal length
\be
(\Delta X)_{min} = \frac{1}{\kappa}\sqrt{ 1 - 2 \frac{\langle |P|\rangle}{\kappa} }
\ee

\subsection{ Deformed momentum representation}
The deformed momentum representation for the algebra (25) is
\be
X = i  \lb 1 - \frac{|p|}{\kappa}\rb^2 \frac{\p}{\p p}
, ~~~
P = p
\ee
The momentum representation acts on the square integrable functions $\Phi(p) \in {\cal L}^2 \lb -\kappa, \kappa; \frac{dp}{\lb 1 - \frac{|p|}{\kappa}\rb^2}\rb$  with $\Phi(\pm \kappa) =0$ and
the norm of $\phi $ is given by
\be
|| \Phi ||^2 = \int_{-\infty}^{\infty}  \frac{ dp}{\lb 1 - \frac{|p|}{\kappa}\rb^2} |\Phi (p)|^2
\ee
The Schr\"odinger equation reads
\be
\left[ \frac{p^2}{2m} + V\lb i  \lb 1 - \frac{|p|}{\kappa}\rb^2 \frac{\p}{\p p} \rb  \right] \Phi(p) = E\Phi(p)
\ee

\subsection{ Position representation}
The position representation for the algebra (25) is
\be
X = x, ~~~~
P =  \frac{p}{ 1 + \frac{|p|}{\kappa}}= \frac{\frac{1}{i}\p_x }{ 1 + \frac{|\frac{1}{i}\p_x|}{\kappa}}
\ee
The position representation acts on the square integrable functions $\psi(x) \in {\cal L}^2 \lb -\infty , \infty; dx \rb  $  and
the norm of $\psi $ is given by
\be
|| \psi ||^2 = \int_{-\infty}^{\infty}  dx  |\psi (x)|^2
\ee
The Schr\"odinger equation reads
\be
\left[ \frac{1}{2m} \left( \frac{\frac{1}{i}\p_x }{ 1 + \frac{|\frac{1}{i}\p_x|}{\kappa}}\right)^2 +  V\lb x\rb  \right] \psi(x) = E\psi(x)
\ee

\subsection{ Undeformed momentum representation}
The undeformed momentum representation for the algebra (25) is obtained from the position representation with replacing $x = i \p_p$, as
\be
X = x = i \p_p, ~~~~
P = \frac{p}{ 1 + \frac{|p|}{\kappa}}
\ee
The undeformed momentum   representation acts on the square integrable functions $\phi(p) \in {\cal L}^2 \lb -\infty , \infty; dp \rb  $  and
the norm of $\phi $ is given by
\be
|| \phi ||^2 = \int_{-\infty}^{\infty}  dp  |\phi (p)|^2
\ee
The Schr\"odinger equation reads
\be
\left[ \frac{1}{2m} \lb \frac{p}{ 1 + \frac{|p|}{\kappa}} \rb^2  + V\lb i \p_p \rb  \right] \phi(p) = E\phi(p)
\ee

\section{ Momentum wave function in a position representation}
Now let us consider the momentum wave function as
\be
P u_{p_0}(x) = p_0 u_{p_0}(x),
\ee
where $p_0$ is not an operator but a momentum eigenvalue. For the exponential function we know
 \be
 f(p)e^{ i ax } =  f\lb \frac{1}{i} \p_x \rb e^{ i ax } = f(a) e^{i ax},
 \ee
Let us assume that the the momentum wave function takes the following form
\be
u_{p_0}(x) = A(p_0) e^{ i a x}
\ee
Inserting the eq.(39) into the eq.(37) we get
\be
P (  A(p_0) e^{ i a x}) = \left[ \frac{p}{ 1 + \frac{|p|}{\kappa}}\right] (  A(p_0) e^{ i a x})
= \left[ \frac{a}{ 1 + \frac{|a|}{\kappa}}\right] (  A(p_0) e^{ i a x})
\ee
Thus, we have
\be
p_0 = \frac{a}{ 1 + \frac{|a|}{\kappa}}
\ee
Taking the absolute value of the eq.(41) we get
\be
|p_0| = \frac{|a|}{ 1 + \frac{|a|}{\kappa}}
\ee
or
\be
|a| = \frac{ |p_0|}{ 1 - \frac{|p_0|}{\kappa}}
\ee
which gives
\be
a = \frac{ p_0}{ 1 - \frac{|p_0|}{\kappa}}
\ee
Thus, we get
\be
u_{p_0}(x) = A(p_0) e^{ i \lb \frac{ p_0 x}{ 1 - \frac{|p_0|}{\kappa}} \rb }
\ee
Using the completeness relation
\be
\int_{-\infty}^{\infty} u_{p_0}(x)^* u_{p'_0}(x) dx = \delta ( p - p'),
\ee
we have
\be
A(p_0) = \frac{1}{ 1 - \frac{|p_0|}{\kappa}}
\ee
Thus, the momentum wave function reads
\be
u_{p_0}(x) = \frac{1}{ 1 - \frac{|p_0|}{\kappa}} \exp \lb \frac{ i p_0x}{ 1 - \frac{|p_0|}{\kappa}}\rb
\ee

\section{One dimensional Box problem in a position representation}

Consider a spinless quantum particle with mass $m$ confined to the following one-dimensional
box
\be
V (x) =
\begin{cases}
0 ~~~ ( 0<x< L)\\
\infty~~~ elsewhere
\end{cases}
\ee
The Schr\"odinger equation in the position representation
reads
\be
\frac{1}{2m} P^2 \psi(x) = E \psi(x)
\ee
or
\be
 \frac{1}{2m} \lb \frac{\frac{1}{i} \p}{ 1 + |\frac{1}{i\kappa} \p|} \rb^2  \psi(x) = E \psi (x)
 \ee
 The solution of the eq.(51) is 
  \be
 \psi (x) =  c_1 \cos \lb \frac{qx}{ 1 - \frac{q}{k}} \rb 
  + c_2 \sin \lb \frac{qx}{ 1 - \frac{q}{k}} \rb 
  \ee
     where
 \be
 q = \sqrt{ 2m E}
 \ee
 From $\psi(0)=0$ we get
 \be
 c_1 = 0
 \ee
 and from $\psi(L)=0$ we get
 \be
  q = q_n = \frac{ n \pi}{ L + \frac{ n\pi}{\kappa}}
   \ee
 where $ n=1, 2, 3, \cdots$. Thus, the wave function is
 \be
 \psi_n (x) = \sqrt{\frac{2}{ L} } \sin \frac{ n \pi}{L} x 
 \ee
  and the energy is given by
 \be
 E_n = \frac{1}{2m} \left[ \frac{ n \pi}{ L + \frac{ n\pi}{\kappa}}\right]^2 
  \ee
  The expectation values of the position and position squared are given by
   \be
   \langle  X \rangle= \frac{L}{2}
   \ee
   \be
   \langle  X^2 \rangle= L^2 \lb \frac{1}{3} - \frac{1}{ 2 n^2 \pi^2} \rb
      \ee
   and the expectation values of the momentum  and  momentum squared are given by
   \be
   \langle  P \rangle=  0
         \ee
   and 
      \be
   \langle  P^2 \rangle= \left[ \frac{ n \pi}{ L + \frac{ n\pi}{\kappa}}\right]^2 
   \ee
   Thus, the uncertainty relation reads
   \be
   \Delta X \Delta P
   = \frac{ n \pi }{ 1 + \frac{n \pi}{ \kappa L}}  \sqrt{ \frac{1}{12} - \frac{1}{2 (n \pi)^2 } }
   \ee
     Fig.1 shows the plot of $E_n$ versus $n$ for $1/\kappa =0$ (brown),  $1/\kappa =0.05$ (green) and  $1/\kappa =0.1$ (pink) with $2m=1, L=\pi$. We know that the energy decreases due to the DSR-GUP effect.

\section{ Harmonic oscillator problem in the undeformed momentum representation}
Let us consider a particle with mass $m$ confined in an harmonic potential
$ V(X) = \frac{1}{2} mw^2 X^2 $. The Schr\"odinger equation reads
\be
\left[ \frac{P^2}{2m} +  \frac{1}{2} mw^2 X^2 \right] \phi = E\phi
\ee
Using the  undeformed momentum representation we get
\be
\left[ \frac{1 }{2m} \lb \frac{p}{1  +\frac{|p|}{\kappa}}\rb^2 - \frac{1}{2} mw^2 \p_p^2  \right] \psi (p)= E\psi (p)
\ee
If we introduce $ s = 1 + |p|/\kappa$, we get
\be
\left[ \frac{d^2}{ds^2} - \frac{\kappa^4}{m^2 w^2} \lb \frac{1}{s^2} -  \frac{2}{s} \rb+ \frac{ 2 \kappa^2 E m- \kappa^4}{( m w)^2} \right]  \psi (s)= 0
\ee
Letting $ s = Ay$, we have
\be
\left[
\frac{d^2}{dy^2} - \frac{\kappa^4}{m^2 w^2 y^2 }+ \frac{2 A \kappa^4}{m^2 w^2 y }+ \frac{ A^2(2 \kappa^2 E m- \kappa^4)}{( m w)^2} \right]  \psi (y)= 0
\ee
Comparing the above equation with the Wittacker equation
\be
\left[
\frac{d^2}{dy^2} + \frac{ \frac{1}{4} - \mu^2 }{y^2 }+ \frac{\lambda}{ y }- \frac{1}{4} \right]  \psi (y)= 0 \ee
we have
\be
A = \frac{mw}{ 2 \kappa^2 \sqrt{ 1 - \frac{2Em}{\kappa^2}}}
\ee
\be
\mu = \frac{1}{2} \sqrt{ 1 +  \frac{4 \kappa^4}{m^2 w^2}}
\ee
\be
\lambda = \frac{\kappa^2}{ m w  \sqrt{ 1 - \frac{2Em}{\kappa^2}}}
\ee
Thus, the wave function in the momentum representation reads
\be
\psi (p) = M_{\lambda, \mu } (y)
\ee
where the  Wittaker function is defined as
\be
 M_{\lambda, \mu } (y) =e^{ -y/2} y^{\mu +1/2} M\lb \mu - \lambda + \frac{1}{2}, 1 + 2 \mu; y\rb
 \ee
 and $M(a,b,y)$ is Kummer function defined by
 \be
  M(a,b,z)=\sum _{{n=0}}^{\infty }{\frac {a^{{(n)}}z^{n}}{b^{{(n)}}n!}}={}_{1}F_{1}(a;b;z)
  \ee
  and  the rising factorial is defined as
  \be
   a^{(0)}=1, ~~~~a^{{(n)}}=a(a+1)(a+2)\cdots (a+n-1)\,,
   \ee
   From the termination of the infinite series, we have the following quantization rule
   \be
   \mu - \lambda + \frac{1}{2} =-n, ~~~~n=0, 1, 2, \cdots
   \ee
   or
   \be
   \frac{1}{2} \sqrt{ 1 +  \frac{4 \kappa^4}{m^2 w^2}} - \frac{\kappa^2}{ m w  \sqrt{ 1 - \frac{2Em}{\kappa^2}}} + \frac{1}{2} =-n
   \ee
   Solving the eq.(76) with respect to $E$ we get
   \be
   E_n = \frac{ mw^2 \kappa^2 \left[
    1 + 2n \lb 1 + n + \sqrt{ 1 + \frac{4 \kappa^4}{ m^2 w^2}}\rb + \sqrt{ 1 + \frac{4 \kappa^4}{ m^2 w^2}}\right]}{
    4 \kappa^4 + 2 m^2 w^2 \left[ 1 + 2n \lb 1 + n + \sqrt{ 1 + \frac{4 \kappa^4}{ m^2 w^2}}\rb + \sqrt{ 1 + \frac{4 \kappa^4}{ m^2 w^2}} \right]}
    \ee
    or
        \be
E_n =  w \lb  n+\frac{1}{2} \rb  - \frac{mw^2}{4 \kappa^2} ( 6n^2 + 6n+1)+ {\cal O} \lb \frac{1}{\kappa^4}\rb \ee
Fig.2 shows the plot of $E_n$ versus $n$ for $1/\kappa =0$ (brown),  $1/\kappa =0.1$ (green) and  $1/\kappa =0.2$ (pink) with $m=1, w=1$. We know that the energy decreases due to the DSR-GUP effect.

\section{Conclusion}

In this paper we found the GUP based on the DSR where the coordinate remain unaltered at the high energy while the momentum is deformed at the high energy so that it may be bounded from the above. Based on the DSR, we considered the relation between the momentum at high energy,  $P$ and the momentum at low  energy,  $p$ as
\be
P = \frac{p}{ 1 + \frac{|p|}{\kappa}},
\ee
which gave the maximum momentum ( Planck momentum). Then, our GUP ( DSR-GUP)
took the following form:
\be
[ X, P] = i \lb 1 - \frac{|P|}{\kappa}\rb^2,
\ee
which gives the minimal length with the maximal momentum ( Planck momentum). Based on the DSR-GUP, we discussed three problems; One dimensional box problem,  momentum wave function, and harmonic oscillator problem. We found that for the one dimensional box problem and harmonic oscillator problem we obtained the exact form of the wave function and energy levels. We also found that for two examples the energy decreases due to the DSR-GUP effect.

It seems interesting to apply the DSR-GUP model to the relativistic quantum mechanics for the purpose of constructing the quantum field theory based on the DSR-GUP. We think that these problems and related topics would be clear in the near future.

\section*{Acknowledgement}
We acknowledges to reviewer for his ( her) helpful comments and encouragement .
This work was supported by the National Research Foundation of Korea Grant funded by the Korean Government (NRF-2015R1D1A1A01057792) and by Development Fund Foundation, Gyeongsang National University, 2018.

\section*{Refernces}

$~~~~$[1] H.S. Snyder, Phys.Rev. 71, 38 (1947)

[2] C.N. Yang, Phys. Rev. 72, 874 (1947)

[3] C.A. Mead,
Phys. Rev. 135, B 849 (1964)

[4] F. Karolyhazy, Nuovo Cim. A 42, 390 (1966).

[5] D. Amati, M. Ciafaloni, G. Veneziano, Phys. Lett. B 197, 81 (1987).

[6] D.J. Gross, P.F. Mende,
8
Phys. Lett. B 197, 129 (1987).

[7] D. Amati, M. Ciafaloni, G. Veneziano, Phys. Lett. B 216,
41 (1989).

[8] K .Konishi, G. Paffuti, P. Provero, Phys. Lett. B 234, 276 (1990).

[9] G. Veneziano,
Proceedings of PASCOS 90, Quantum Gravity near the Planck scale, Boston 1990, p.486
(unpublished).

[10] S. Capozziello, G. Lambiase, G. Scarpetta, Int. J. Theor. Phys. 39, 15 (2000).

[11] M. Maggiore, Phys. Lett. B 304, 65 (1993).

[12] A. Kempf, G. Mangano, R.B. Mann, Phys. Rev. D 52, 1108 (1995).

[13]  M. Bojowald, A. Kempf,
Phys. Rev. D 86, 085017 (2012).

[14] F. Scardigli, Phys. Lett. B 452, 39 (1999).

[15] R.J. Adler, D.I. Santiago, Mod. Phys. Lett. A14, 1371 (1999).

[16] F. Scardigli and R. Casadio, Class. Quantum Grav. 20, 3915 (2003).

[17] P. Pedram, Phys. Rev. D 91 (2015) 063517.

[18]  P. Pedram,  Adv. High Energy Phys. 2013, 853696 (2013)

[19]  P. Pedram, Adv. High Energy Phys. 2016 (2016) 5101389

[20]  P. Pedram, Europhys. Lett. 101 (2013) 30005

[21] P. Pedram, J. Phys. A: Math. Theor. 45 (2012) 505304

[22] P. Pedram,  Int. J. Theor. Phys. 51 (2012) 1901-1910

[23] P. Pedram,  Phys. Lett. B 714 (2012) 317

[24] P. Pedram,  Phys. Lett. B 718 (2012) 638

[25] P. Pedram,  Int.J.Mod.Phys.D19:2003-2009,2010

[26] P. Pedram, K.Nozari, S. H. Taheri, JHEP 1103:093,2011

[27] K. Nozari, P. Pedram,  EPL 92 (2010) 50013

[28] K. , M. Moafi, F. Rezaee Balef, Advances in High Energy Physics 2013 (2013) 252178

[29] M. Asghari, P. Pedram, K. Nozari, Phys. Lett. B 725, 451 (2013)

[30] J. Vahedi, K. Nozari, P. Pedram, Gravitation and Cosmology 18, 211 (2012)

[31] K. Nozari, A. Etemadi, Phys. Rev. D 85 (2012) 104029

[32] K. Nozari, P. Pedram, M. Molkara ,  Int. J. Theor. Phys. 51 (2012) 1268-1275

[33] H.Shababi, Won Sang Chung, Phys.Lett.B 770 (2017) 445.

[34] K.Nouicer, Phys.Lett.B646 (2007) 63.

[35] H. Shababi, P. Pedram and Won Sang Chung, Int.J.Mod.Phys.A31 (2016) 1650101.

[36] G. Amelino-Camelia, Int. J. Mod. Phys. D11, 35 (2002).

[37] G. Amelino-Camelia, Phys. Lett.
B510, 255 (2001).

[38] J. Lukierski, A. Nowicki, H. Ruegg and V.N. Tolstoy, Phys. Lett. B264,
331 (1991).

[39] J. Lukierski, A. Nowicki and H. Ruegg, Phys. Lett. B293,
344 (1992).

[40] J. Magueijo and L. Smolin, Phys. Rev. Lett. 88, 190403 (2002).
\newpage

\begin{figure}
\includegraphics[width=8cm]{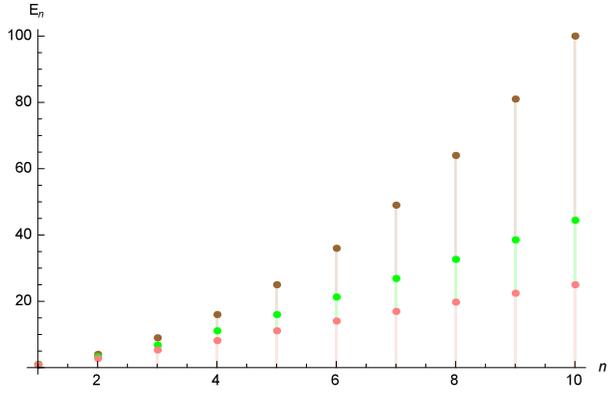}
\caption{plot of $E_n$ versus $n$ for $1/\kappa =0$ (brown),  $1/\kappa =0.05$ (green) and  $1/\kappa =0.1$ (pink) with $2m=1, L=\pi$.
 }
\label{ex1}
\end{figure}

\begin{figure}
\includegraphics[width=8cm]{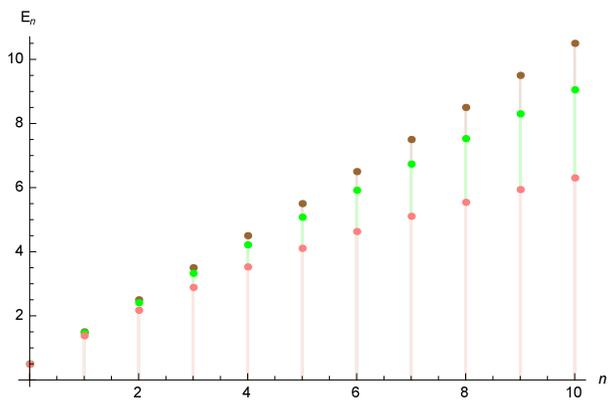}
\caption{plot of $E_n$ versus $n$ for $1/\kappa =0$ (brown),  $1/\kappa =0.1$ (green) and  $1/\kappa =0.2$ (pink) with $m=1, w=1$.
 }
\label{ex1}
\end{figure}

\end{document}